\documentclass[mathleft
]{an}
\usepackage{graphicx}
\usepackage{times}
\usepackage{subfigure}
\def\lta{\mathrel{\hbox{\rlap{\hbox{\lower4pt\hbox{$\sim$}}}\hbox{$<$}}}}
\def\gta{\mathrel{\hbox{\rlap{\hbox{\lower4pt\hbox{$\sim$}}}\hbox{$>$}}}}

\begin{document}

\Pagespan{0}{0}
\Yearpublication{2012}%
\Yearsubmission{2012}%
 \DOI{XX}%

\title{Solving the Cooling Flow Problem through Mechanical AGN Feedback}

\author{M. Gaspari\inst{1}\fnmsep\thanks{Corresponding author: massimo.gaspari4@unibo.it\newline},
F. Brighenti\inst{1},
\and  
M. Ruszkowski\inst{2}
}
\titlerunning{Solving the Cooling Flow Problem through Mechanical AGN Feedback}
\authorrunning{Gaspari, Brighenti \& Ruszkowski}
\institute{Department of Astronomy, University of Bologna, Via Ranzani 1, 40127 Bologna, Italy
\and 
Department of Astronomy, University of Michigan, 500 Church Street, Ann Arbor, MI 48109, USA}

\received{22 Aug 2012}
\accepted{13 Sep 2012}


\abstract{Unopposed radiative cooling of plasma would lead to the cooling catastrophe, a massive inflow of condensing gas, manifest in the core of galaxies, groups and clusters. The last generation X-ray telescopes, {\it Chandra} and {\it XMM}, have radically changed our view on baryons, indicating AGN heating as the balancing counterpart of cooling. This work reviews our extensive investigation on self-regulated heating. We argue that the mechanical feedback, based on massive subrelativistic outflows, is the key to solving the cooling flow problem, i.e. dramatically quenching the cooling rates for several Gyr without destroying the cool-core structure. Using a modified version of the 3D hydrocode FLASH, we show that bipolar AGN outflows can further reproduce fundamental observed features, such as buoyant bubbles, weak shocks, metals dredge-up, and turbulence. The latter is an essential ingredient to drive nonlinear thermal instabilities, which cause the formation of extended cold gas, a residual of the quenched cooling flow and, later, fuel for the feedback engine. Compared to clusters, groups and galaxies require a gentler mechanical feedback, in order to avoid catastrophic overheating. 
We highlight the essential characteristics for a realistic AGN feedback, with emphasis on observational consistency.}

\maketitle

\section{The cooling flow problem}
A fundamental gap in the current understanding of galaxies, groups and clusters concerns the thermodynamical
evolution of the baryonic component, mainly formed by the extended halo of diffuse plasma ($10^7-10^8$ K).
Emitting strongly X-ray radiation, the central hot gas would lose most of its thermal energy and pressure support
in $10-100$ Myr. The final result would be a massive cooling flow, initiating from the denser central regions and inducing the peripheral gas (up to 100s kpc) to flow subsonically toward the nucleus. The cooling rates predicted by the classic theories are unrealistic, reaching $1000$s $M_{\odot}$ yr$^{-1}$ (Fabian 1994). 
In the core, huge amount of cool gas (and stars) would rapidly accumulate, monolithically condensing out of the hot phase and creating an unobserved large peak in surface brightness. 

This is the {\it cooling flow problem}.

\section{The AGN heating solution}

The amazing details provided by the last-generation X-ray telescopes, {\it Chandra} and {\it XMM-Newton},
have revealed that the active galactic nucleus (AGN), locus of a supermassive black hole (SMBH),
strongly interacts with the surrounding medium in the form of bubbles, jets, outflows, shocks,
sonic ripples, turbulence and gas entrainment (a striking example is NGC 5813; Randall et al. 2011). The inferred energies released by the AGN (up to $10^{62}$ erg) are indeed capable to balance the radiative losses. 
We now need to find
the ultimate origin of those features and, especially, how the AGN energy couples to the surrounding gas. 

\begin{figure*}
        \center
        \subfigure{\includegraphics[scale=0.60625]{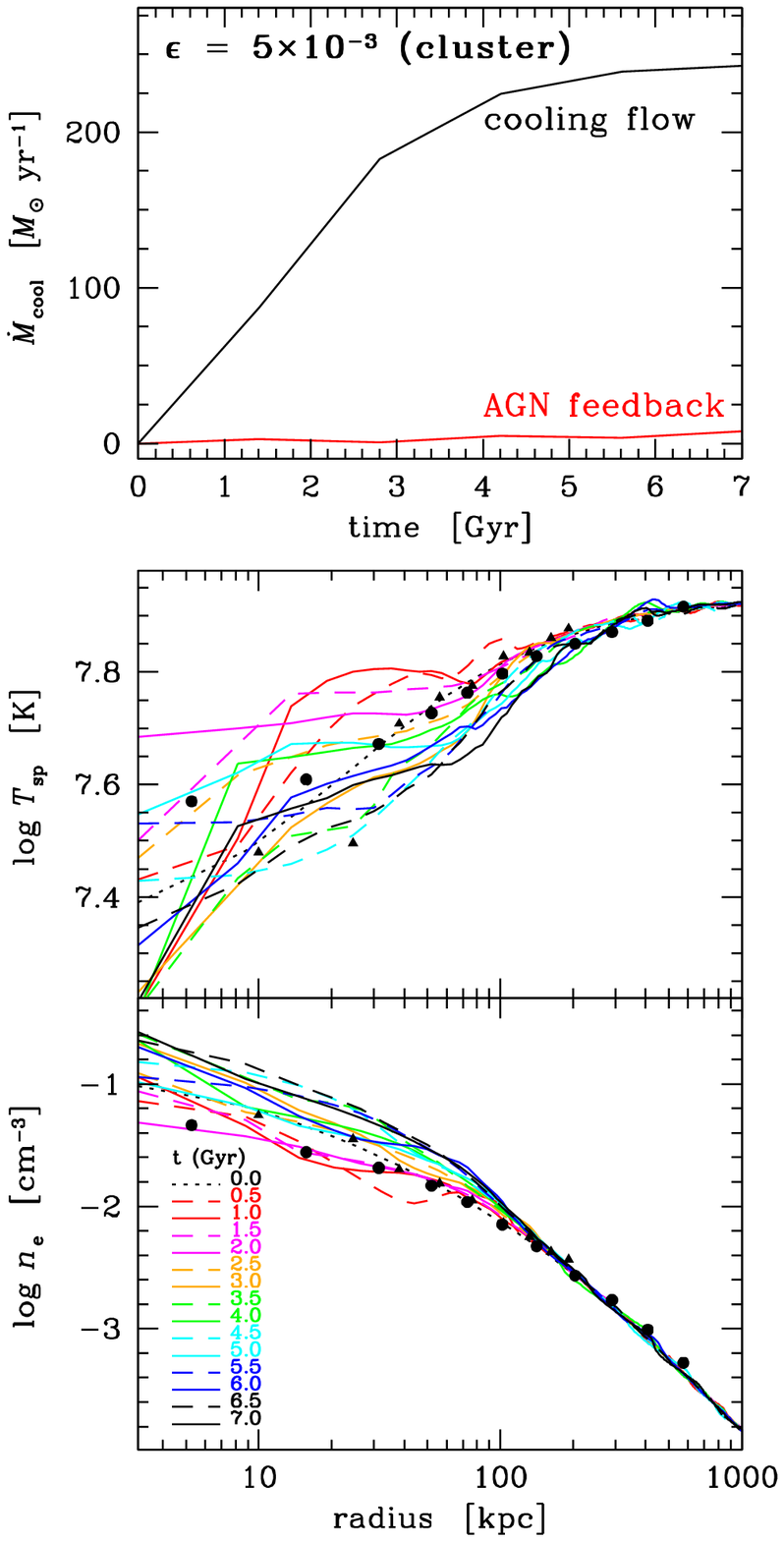}}
        \subfigure{\includegraphics[scale=0.60625]{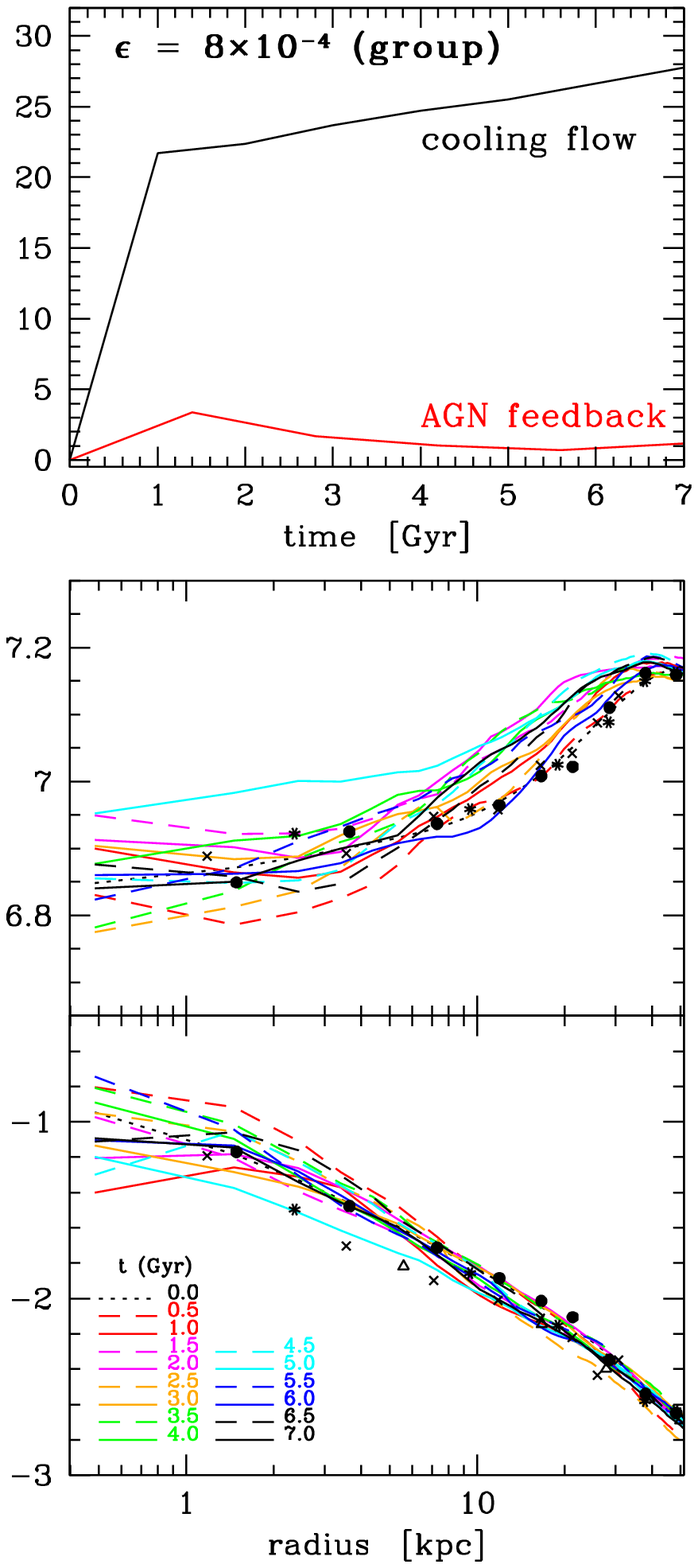}}
        \subfigure{\includegraphics[scale=0.60625]{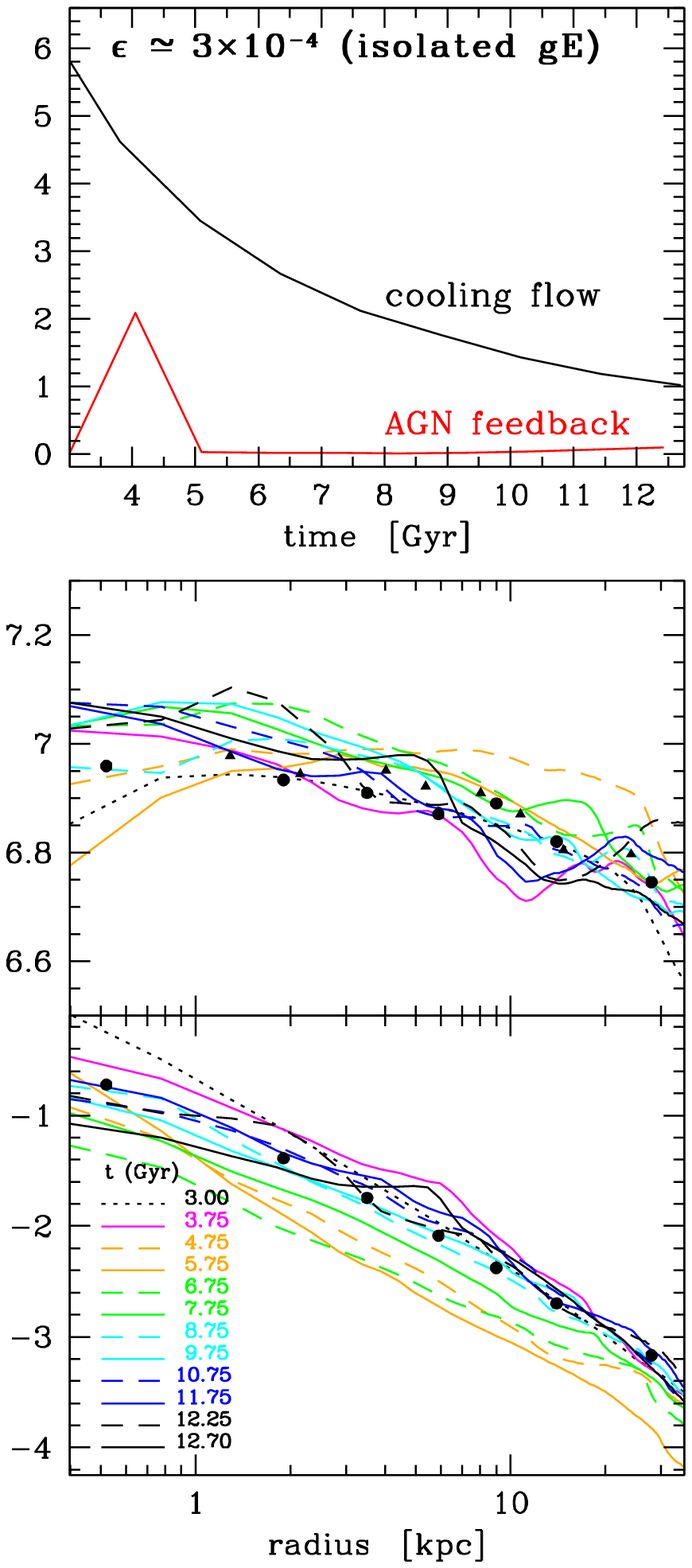}}
        \caption{\small{Exemplary simulations of self-regulated mechanical AGN outflows in a galaxy cluster (A1795), group (NGC 5044), and isolated giant elliptical galaxy (NGC 6482). First row: cooling rates in function of time; note the steady order of magnitude suppression due to AGN feedback. Second and third row: radial profiles of spectroscopic-like temperature and electron number density; the mechanical feedback is able to consistently preserve the cool-core structure, i.e. avoiding for several Gyr the overheating and emptying of the central region. Revised and adapted from Gaspari et al. 2011a, 2012b (refer to these articles for further details).}}
          \label{fig:1}
\end{figure*}  

The leitmotiv of our previous works (Gaspari et al. 2009, 2011a,b, 2012a,b -- G11a,b, G12a,b) is that massive subrelativistic AGN outflows are the key to solving the cooling flow problem. Their origin near the BH
is still unclear, either resulting from an entrained and decelerated magnetic jet (e.g. Giovannini 2004, Croston 2008), 
or directly from a nuclear radiative disc wind (Crenshaw et al. 2003).
Nevertheless, on kpc scales the observations point toward a feedback process governed by 
directional mechanical input of energy, with subrelativistic velocities, $\sim10^3-10^4$ km s$^{-1}$, and substantial mass outflow rates $\sim10^{-1}-10^2$  $M_{\odot}$ yr$^{-1}$ (e.g. Nesvadba et al. 2008, Tombesi et al. 2012). 

In the present work, we review the key features for a feedback heating to succeed in quenching the cooling flow.
To be as realistic as possible, we carried out large 3D hydrodynamic simulations, via the state-of-the-art AMR parallel code FLASH, substantially modified to study 
AGN outflows, strong shocks, radiative cooling, thermal instabilities, and multiphase gas. We studied the effects of self-regulated AGN outflows in a wide range of virialised systems, from massive clusters to groups and isolated ellipticals ($M_{\rm vir}\sim10^{13}-10^{15}\ M_{\odot}$), with a maximum resolution of $\sim150-500$ pc and an evolution $\gta7$ Gyr. 
The numerical and physical details are
extensively covered in G11a,b and G12a,b. 
We found that a realistic feedback mechanism should be
primarily mechanical, anisotropic, driven by subrelativistic outflows, and self-regulated by cold gas accretion.
The next Sections highlight the reasons behind that.

\section{Avoid overcooling \& overheating}
The necessary, but not sufficient, condition to assess that the cooling flow problem has been solved is the drastic suppression of the cooling rates, below $5-10\%$ of the classic predictions, consistent with the spectroscopic constraints (Peterson \& Fabian 2006).
In Figure \ref{fig:1} are presented the three exemplary AGN feedback simulations for the galaxy cluster, group and isolated giant elliptical (G11a, G12b). The top row clearly shows the steady quenching of $\dot{M}_{\rm cool}$ by at least an order of magnitude (red), for $\gta7$ Gyr, compared to the pure cooling flow run (black). 

At the same time, any consistent feedback should {\it not erase the cool-core} structure, an often unwelcome -- and ignored -- product of strong concentrated heating (as thermal and quasar-like radiative feedback). Cool cores are in fact ubiquitous among clusters and groups (Vikhlinin et al. 2006; Sun et al 2009). As shown in Figure \ref{fig:1}, second row, the spectroscopic-like 
temperature profiles of our self-regulated feedback models 
maintain the positive moderate gradient typically observed in clusters and groups. 
Due to the higher relevance of compressional heating, the isolated elliptical shows an almost flat temperature, also in the pure cooling run; again, the gentle mechanical feedback consistently prevents the creation of gradient inversions.
Avoiding overheating means also to not totally empty the galaxy for several Gyr (see density radial profiles, $n_{\rm e}$, bottom row in Fig. \ref{fig:1}).
On the other hand, strong density cusps are also avoided, solving the problem of the drastic central {\it SB}$_{\rm X}$ enhancement in cooling flows.

{\it Avoiding overcooling and overheating at the same time}, in a state of quasi thermal equilibrium on large spatial and temporal scales, is crucial. We live now probably in a theoretical {\it era of heating catastrophe},
in which cooling can be easily halted, but models are rarely checked against overheating.
One of the fundamental characteristics of mechanical feedback is that, although outflow events are very powerful ($\sim10^{44}-10^{46}$ erg s$^{-1}$), the injected energy is only {\it gradually} thermalised along the jet path, thus preserving the long-term cool-core structure.

\section{Scaling with the halo mass}
Groups and ellipticals are not scaled-down versions of clusters.
To avoid drastic overheating and erasing the cool core, 
less bound objects require on average a less efficient AGN feedback. 
The mechanical efficiencies of the best (cold accretion) models are (Fig. \ref{fig:1}):
$\epsilon \sim5\times10^{-3}-10^{-2}$ (cluster), $\epsilon\sim5\times10^{-4}-10^{-3}$ (group) and
$\epsilon\sim10^{-4}-5\times10^{-4}$ (gE). 
This $\epsilon$ value does not only represent the micro-scale mechanical efficiency, near
few BH radii, but it englobes the large-scale coupling between the outflows and the environment.
Numerical resolution is also relevant, but convergence tests suggest
that the trend of the decreasing efficiency could be real (G11b, G12b).
As a result of the higher $\epsilon$ and larger accreted mass, the AGN energy injected in the cluster to
quench cooling is much higher (few $10^{62}$ erg),
compared to the group ($\sim10^{61}$ erg) or gE ($\sim10^{60}$ erg). 

Even if rescaled,
the same self-regulated AGN feedback in less bound halos has still
more profound consequences on the core long-term structure, more slowly recovering from stronger outbursts
(e.g. see the isolated gE evolution).
This can contribute to the (observed) break in the self-similar scaling relations at lower masses (e.g. Sun et al. 2009).

Besides the cool-core survival, the physical reason for the mechanical efficiency to be linked to the 
environment/potential well 
needs to be clarified with future works. The fact that the most massive BHs 
reside at the centre of clusters (McConnell et al. 2011) should play a crucial role. 

\section{Feedback imprints}

\subsection{Buoyant bubbles}
The anisotropic injection of mechanical energy is able to naturally inflate pairs of underdense cavities
in the surrounding hot atmosphere (cf. McNamara \& Nulsen 2007).
In the cluster regime, the X-ray cavities have usually a radius of tens kpc, with $T_{\rm X}$ slightly hotter than the ambient medium (in the early stage of inflation). The reduced power in groups and ellipticals produces more gentle cavities (with radius $\lta10$ kpc), showing relatively cold metal-rich rims and mild internal $T_{\rm X}$. The jump in {\it  SB}$_{\rm X}$ is typically $20-40$\%. Figure \ref{fig:2} (top left) depicts a typical example.
The jet-inflated bubbles are also stable for tens Myr (due to the internal vortexes), 
contrary to artificially ad-hoc inflated cavities, more prone to Rayleigh-Taylor instability.

\subsection{Weak shocks}

\begin{figure*}
        \center
        \subfigure{\includegraphics*[scale=0.51]{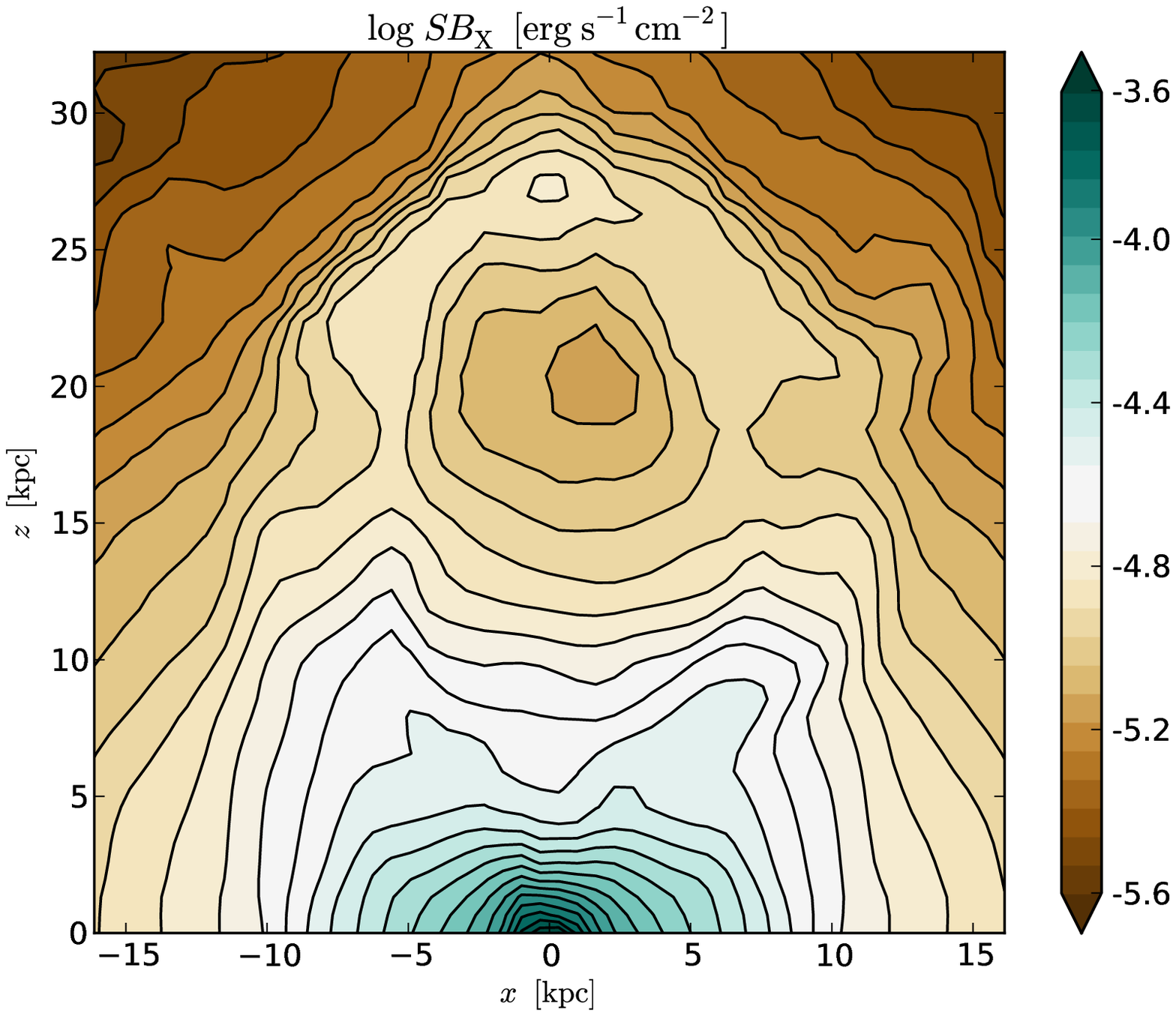}} 
        \subfigure{\includegraphics*[scale=0.51]{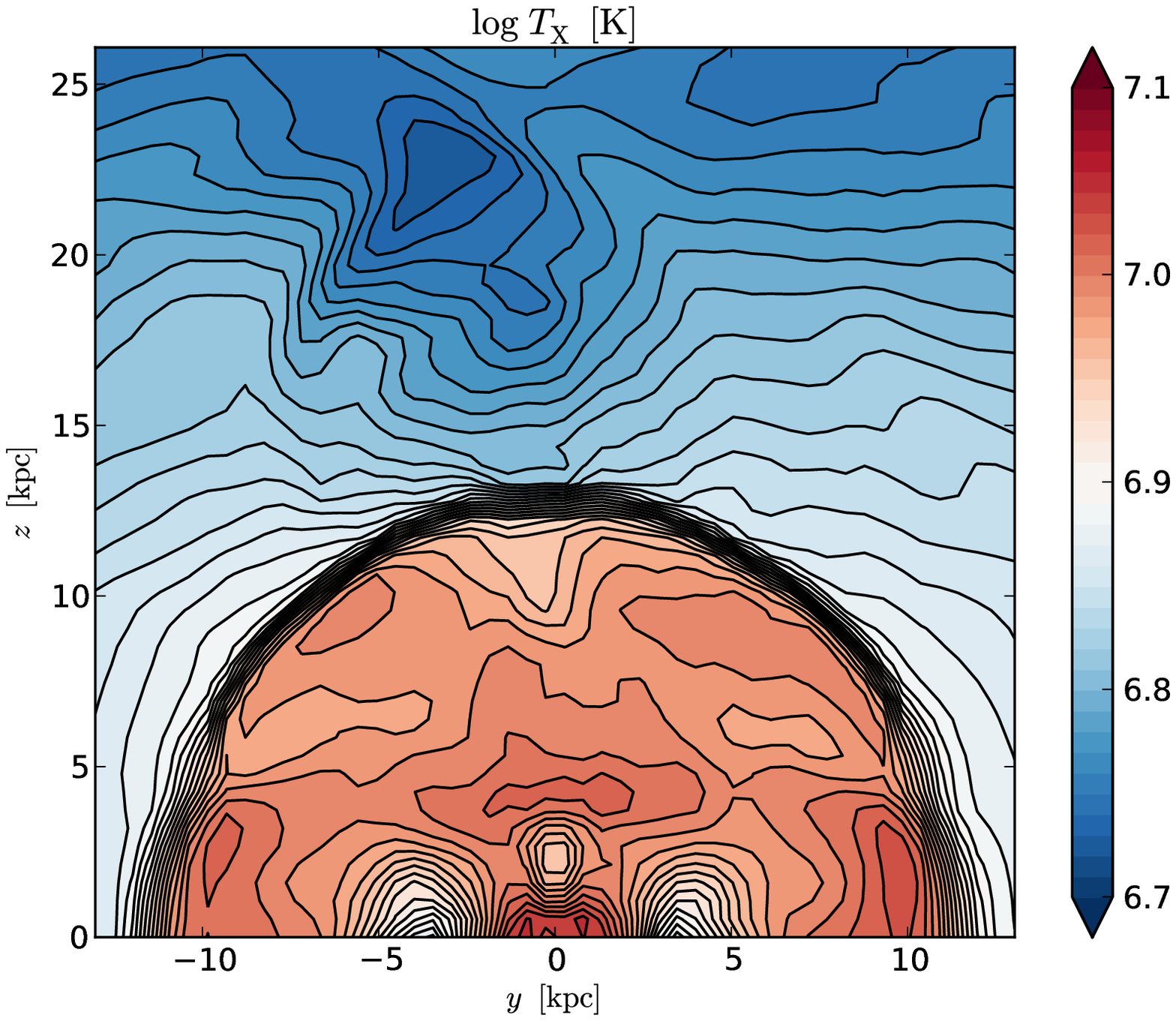}}
        \subfigure{\includegraphics*[scale=0.51]{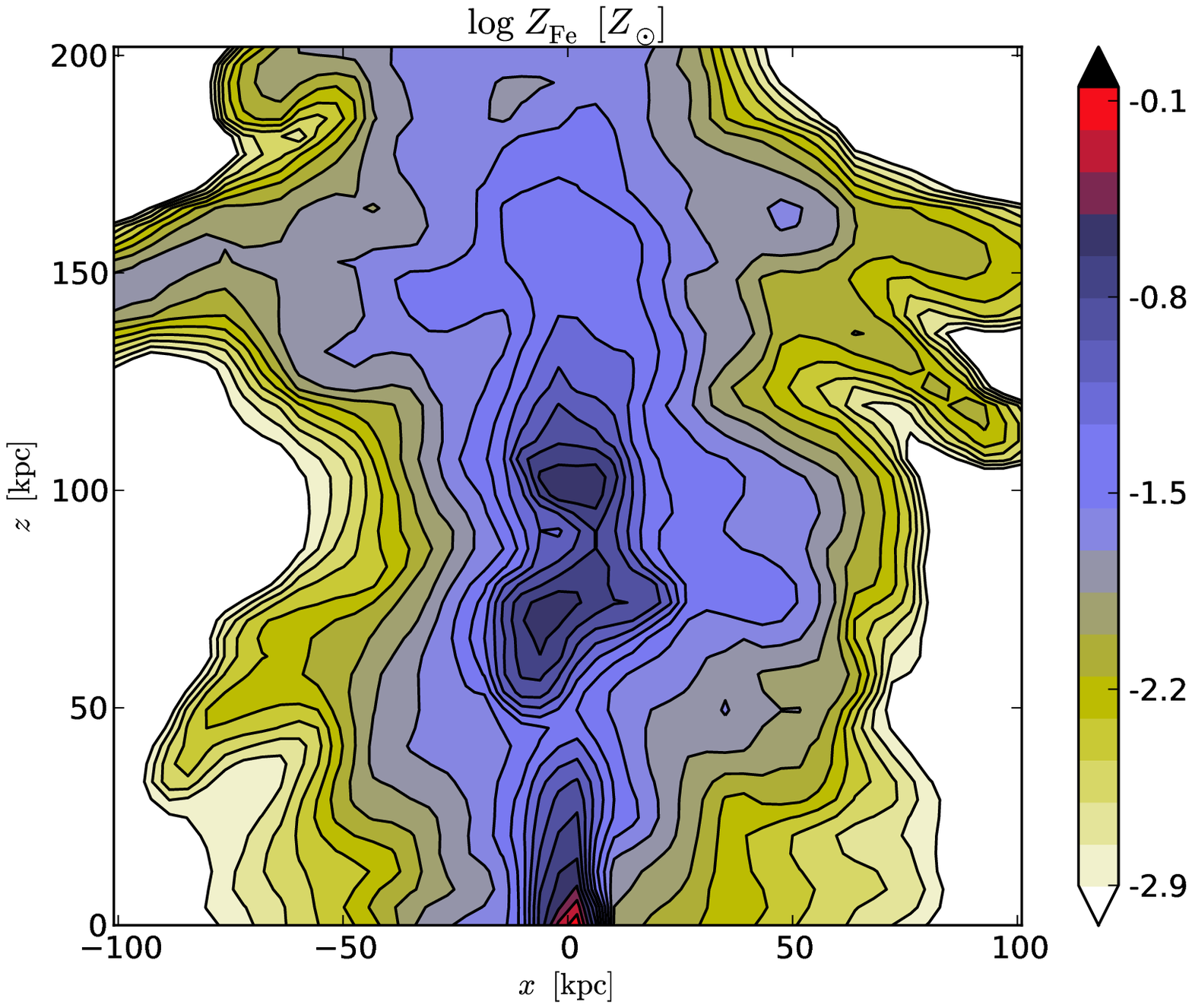}}
        \subfigure{\includegraphics*[scale=0.51]{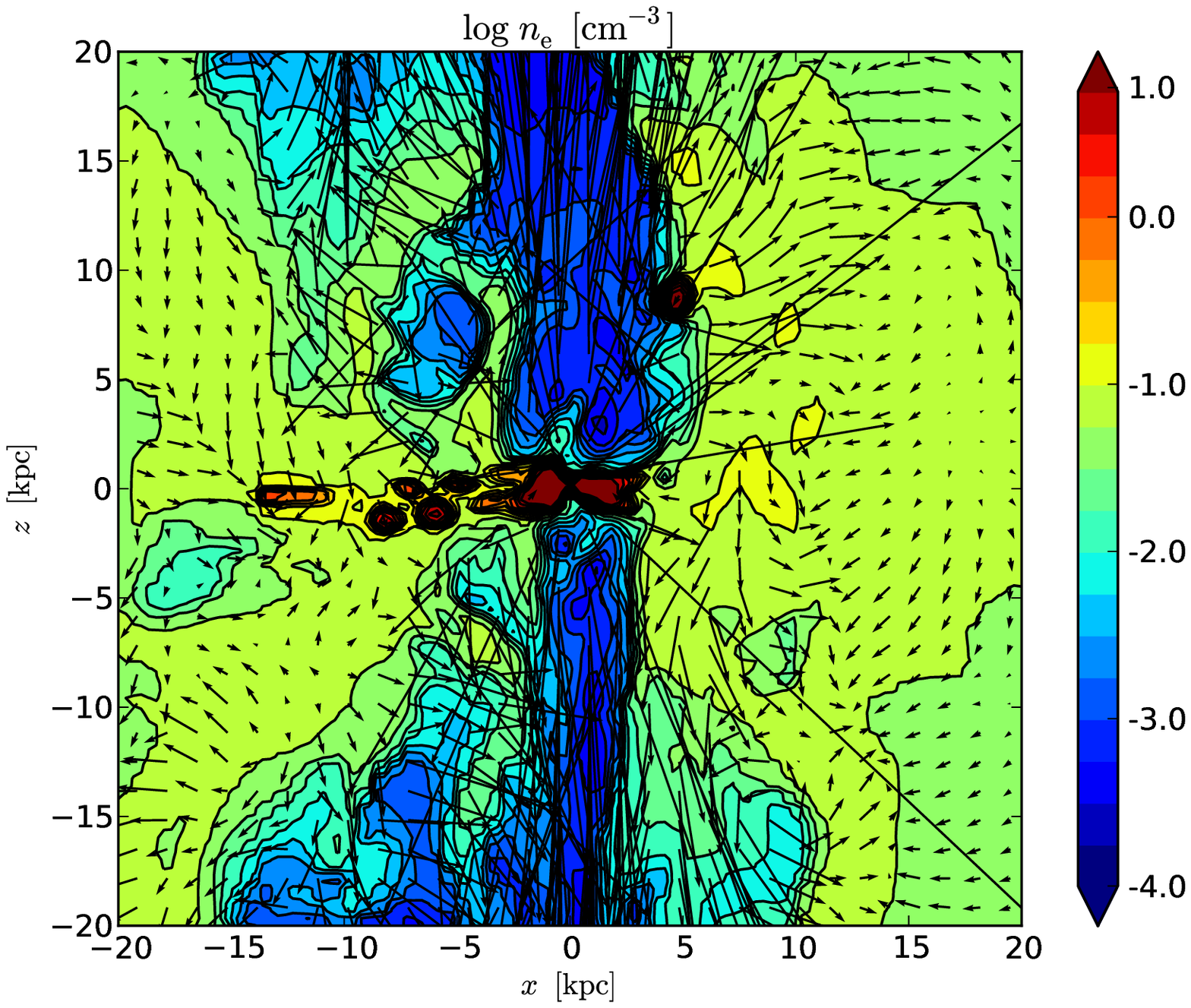}}
        \caption{\small{Essential imprints of the AGN outflow feedback. Top left: 
        underdense cavity (X-ray surface brightness; group). Top right: cocoon shock with ${\rm Mach}\sim1.3$ (projected X-ray temperature; isolated gE). Bottom left: iron dredge-up along jet axis (projected emission-weighted abundance; cluster). Bottom right: turbulence, thermal instabilities and multiphase gas in a cluster core (density cut, with the velocity field overlaid -- an axis bin corresponds to 750 km s$^{-1}$). Revised and adapted from Gaspari et al. 2011a,b, 2012a,b.}}
        \label{fig:2}  
\end{figure*}

The injected mechanical energy is {\it not entirely} transformed into the bubble enthalpy, as often assumed.
At the beginning of the outburst, massive outflows spend most of their energy to generate
the elliptical cocoon shock (with jumps in {\it  SB}$_{\rm X}$ and $T_{\rm X}$; e.g. top right panel in Fig. \ref{fig:2}), enveloping the subsequent bubble. The initial strong shock (${\rm Mach}\sim 5-10$) is statistically unlikely to be observed, 
due to the rapid deceleration. It is thus not surprising that the estimated Mach numbers (e.g. Randall et al. 2011) are often slightly supersonic, $\sim1.1-1.7$, with few exceptions (Centaurus A, Mach$\,\sim8$; NGC 3801, Mach$\,\sim4$). 
The recurrent shocks become extremely weak at larger radii ($\gta50$ kpc), degrading into faint sonic ripples (e.g. Perseus cluster or Fig. \ref{fig:1}).

\subsection{Metal-rich and cold gas dredge-up}
The anisotropic outflow ram pressure is fundamental for uplifting the metals (mainly iron), from the central reservoir,
replenished by the BCG stellar evolution, up to $50-200$ kpc along the jet axis (Figure \ref{fig:2}, bottom left). The typical contrast
with the background is $\sim10-20\%$. This dredge-up is consistent with recent observations (e.g. Kirkpatrick et al. 2009).
During the more quiescent phases, the stirring motions tend to restore the homogeneous metal distribution.
In addition, stronger outbursts can also uplift the central condensed cold gas, creating extended filamentary structures.

\subsection{Outflow-driven turbulence}
Cyclical AGN outflows generate substantial level of turbulence in the core, on scales of $\sim5-15$ kpc (Figure \ref{fig:2}, bottom right). The (emission-weighted) velocity dispersions are usually subsonic, in the range $200-400$ km s$^{-1}$ for the group/elliptical, and few hundred more in the cluster, in line with recent constraints (dePlaa et al. 2012). 
The stronger stirring, due to powerful outbursts and in part compressive,
can substantially alter hydrostatic equilibrium, leading to an error
in the total mass estimate up to a factor of two; the equilibrium is kept instead below $10\%$ during the common moderate phase. The AGN turbulence (growing after 100s Myr) is a key ingredient for the deposition and isotropisation of the mechanical energy in the core, via the fragmentation of the jet channel and the turbulent energy dissipation.

\section{Multiphase gas: by-product \& fuel}
Outflow-driven turbulence is also the ultimate cause of the nonlinear growth of thermal instabilities in the core. 
During phases of slight cooling dominance, the ratio of the cooling and the free-fall time falls below 10 (linked to the entropy threshold of $30$ keV cm$^2$), and dense cold -- $10^4$ K -- gas condenses out of the hot phase. The cold gas morphology is bimodal (Fig. \ref{fig:2}, bottom right):
the extended phase ($r\lta20$ kpc),
in the form of clouds/filaments generated by thermal instabilities
(and occasionally due to jet uplift), 
plus a more spatially concentrated phase,
in the form of a rotating torus due to nuclear cooling or residual clouds (similar to H$\alpha$ surveys, e.g. McDonald et al. 2011).

The extended gas moves then in free-fall toward the SMBH, boosting the accretion rate and promoting the next phase
of slight heating dominance, in a natural cycle of steady quasi thermal equilibrium (Section 3). 
In the end, multiphase gas 
is not only direct {\it consequence} of feedback, but also the fundamental {\it source} of the AGN heating engine. 
\newpage

\acknowledgements
We acknowledge the NASA awards SMD-11-2209, SMD-11-2507, SMD-12-3033 (Pleiades), and
the CINECA awards HP10BPTM62, HP10BOB5U6 (SP6). The FLASH code was in part developed by the DOE NNSA-ASC OASCR Flash Center at the University of Chicago. 



\end{document}